\documentclass[11pt]{article}\hfuzz=10pt  \topmargin=-0.5cm\hfuzz=10pt  \oddsidemargin=0.1cm\textheight=220mm \textwidth=16.5cm
\usepackage{amssymb}
\usepackage{epsfig}
\newcommand{\comm}[1]{ }

\newtheorem{theorem}{Theorem}
\newtheorem{lemma}{Lemma}
\newtheorem{remark}{Remark}
  
\def\defi{\stackrel{{\scriptscriptstyle \Delta}}{=}}

\def\a{\alpha}

\def\w{\widehat}

\def\Re{{\rm Re\,}}

\def\R{{\bf R}}

\def\H{{\cal H}}

\def\L{L}

\def\g{\gamma}
\def\C{{\bf C}}
\def\W{{\cal W}^*}

\def\oo{\bar}

\def\V{{\cal V}}
\def\A{{\cal A}}

\def\L{{\cal L}}

\newcommand{\be}{\begin{equation}}
\newcommand{\ee}{\end{equation}}
\newcommand{\bd}{\begin{displaymath}}
\newcommand{\ed}{\end{displaymath}}
\newcommand{\ba}{\begin{array}{ll}}
\newcommand{\ea}{\end{array}}
\newcommand{\baa}{\begin{eqnarray}}
\newcommand{\eaa}{\end{eqnarray}}
\newcommand{\baaa}{\begin{eqnarray*}}
\newcommand{\eaaa}{\end{eqnarray*}}

\def\L{{\cal L}}

\def\W{{\cal W}}

\def\H{{\cal H}}

\def\R{{\bf R}}

\def\H{H_{BC}}
\def\H{{H^1}}
\def\HH{{H^2}}
\def\HBC{{H_{BC}}}
\title{On recovering of solutions of Schr\"odinger equations from their time averages}
\author{
Nikolai Dokuchaev}
\begin{document}
\maketitle
\let\thefootnote\relax
\footnote{
 }
\let\thefootnote\relax\footnote{The author is with  School of Electrical Engineering, Computing and Mathematical Sciences, Curtin
University,   GPO Box U1987, Perth, 6845 Western Australia. Email
N.Dokuchaev@curtin.edu.au. Ph. 61 8 92663144}
\begin{abstract}
The paper study a possibility to recover  solutions of  Schr\"odinger equations 
from its time-averages in the setting where the values at the initial time are  unknown.  This problem
 can be reformulated as a new boundary value problem where a Cauchy condition is replaced by
 a prescribed time-average of the solution. It is shown that this new problem is well-posed in certain classes of solutions.
  The paper establishes existence, uniqueness,
 and  a regularity of the solution for this new problem.
\\
PACS 2010: 02.30.Zz,    
02.30.Jr.   
\end{abstract}
\section{Introduction}
The most common type of boundary condtitons  for
evolution partial differential equations are Cauchy initial conditions.
It is know that these conditions can be replaced, in some cases, by non-local conditions, for example, including integrals over time intervals.
For hyperbolic equations, some related results and the references   can be found, e.g., in \cite{KP,PS}.
For parabolic equations, some results the references  can be found in \cite{D19,Pri}.  
 For Schr\"odinger equations, some  problems with non-local in time boundary conditions have also  been considered;  see, e.g., \cite{AS,BP,SM}, and the references therein.

The paper readdresses the problem of  solvability  of boundary value problems
with non-local in time conditions for Schr\"odinger equations.  The underlying  problem
 is formulated as a new boundary value problem where a Cauchy condition is replaced by
 a prescribed weighted time-average of the solution.   It is shown that this new problem is well-posed in a wide enough  class of solutions. 

 Related results were obtained   In \cite{AS,BP,SM} for Schr\"odinger equations with   non-local in time conditions.
   In  \cite{AS,SM}, the nonlocal conditions connected solutions in a finite set of times.  In \cite{BP}, the conditions were quite general and  allowed to include integrals over time.
 The method used in \cite{AS,BP,SM}   was based on the contraction mapping theorem; this
 allowed to cover conditions that can be represented as $u(0)-B(u|_{t\in[0,T]})=\mu$, where $\mu$ is the free term,  $u=u(t)$ is the solution, $t>0$ is the time, and $B$ is an operator  with sufficiently small norm such that the contraction mapping theorem can be applied.
 
 In the present method, we use a different method based the spectral expansion, similarly to the setting from \cite{D19,GG,LV}. This allowed to consider non-local conditions
 where $u(0)$ is not presented as a dominating term, such as conditions
 $\int_0^Te^{rt}u(t)dt=\mu$, for some number $r$.  
 We  establishe existence, uniqueness,
 and  a regularity of the solution for this new problem.
In addition, the eigenfunction expansion for the solution is presented explicitly. 
This allows  to derive a numerical solution. 
 
 The rest of the paper  is organized as follows. In Section \ref{SecS}, we introduce
boundary value problem with averaging over time.
In Section \ref{SecM}, we present the main result  (Theorem \ref{ThM}).
In Section \ref{SecP}, we present the proofs.

\section{Some definitions}\label{SecS}
For a Banach space $X$,
we denote  the norm by $\|\cdot\|_{ X}$. For a Hilbert space $X$,
we denote  the inner product by $(\cdot,\cdot)_{ X}$. We denote
 the Lebesgue measure and
 the $\sigma $-algebra of Lebesgue sets in $\R^n$
by $\oo\ell _{n}$ and $ {\oo{\cal B}}_{n}$, respectively.

 Let $D\subset \R^n$ be a domain, and let $H=L_2(D,\oo{\cal B}_n, \oo\ell_{1}; \C)$ be the space of complex-valued functions.  Let $A:H\to H$ be a self-adjoint operator defined
 on an everywhere closed  subset $D(A)$ of $H$ such that if $u$ is a real valued function then $Au$ is also a real valued function.

 Let $\HBC$ be a set functions from $H$ such that $\HBC$ is everywhere dense in $H$, that the set $D(A)\cap \w H$ is everywhere dense in  $\w H$, and that $(u, Av)_H$ is finite and defined for $u,v\in \HBC$ as a continuous extension from $D(A)\times D(A)$.

Consider an eigenvalue problem \baa
 Av=-\lambda v,\quad v\in\HBC.  \label{EP}\eaa
We assume that this equation is satisfied for $v\in \HBC$  if
  \baaa
 (w,Av)_H=-\lambda (w,v,)_H \quad \forall w\in\HBC  \label{EPw}.\eaaa

Assume that  there exists
a  basis  $\{v_k\}_{k=1}^\infty\subset \HBC$  in $H$  such that
 $$(v_k,v_m)_{H}=0,\quad  k\neq m,\quad \|v_k\|_{H}=1,$$
and that $v_k$ are eigenfunctions for (\ref{EP}), i.e.,
\baaa
 Av_k=-\lambda_k v_k, \label{EPk}\eaaa
for some $\lambda_k\in\R$ such that $\inf_k\lambda_k>-\infty$, $k\to +\infty$.

These assumptions imply that the operator $A$ is self-adjoint with respect to the boundary conditions  defined by the choice of $\HBC$. 

Let $\HH$ be the set of functions from $\HH$ such that  $(Au, Au)_H$ is finite.

It follows from the assumptions that there exists $c_A\ge 0$  such that
 \baaa
  &&-(u, Au)_H+c_A\|u\|_H^2> 0,\quad u\neq 0,\quad u\in \H,
\\
 &&(Au, Au)_H+c_A\|u\|_H^2> 0,\quad u\neq 0,\quad u\in \HH.
 \eaaa

We consider $H$ and $\HH$ as Hilbert spaces 
with the  norms  \baaa
&&\|u\|_{\H}^2\defi (u, Au)_H+c_A\|u\|_H^2,\\
&&\|u\|_{\HH}^2\defi (Au, Au)_H+c_A\|u\|_H^2.
 \eaaa

Clearly, the spaces $\H$ and $\HH$ are isometric to  weighted $\ell_2$-spaces: if $u =\sum_{k=1}^\infty c_k v_k$, then
 \baaa
\|u\|_{\H}^2=\sum_{k=1}^\infty (\lambda_k+c_A)|c_k|^2,\quad \|u\|_{\HH}^2=\sum_{k=1}^\infty (\lambda_k^2+c_A)|c_k|^2.
 \eaaa
\par
Introduce the spaces
\baaa
&&{\cal C}\defi C\left([0,T]; H\right),\quad {\cal C}^k\defi C\left([0,T]; H^k\right),\quad k=1,2,
\quad \W\defi
L^{2}\bigl([ 0,T ],\oo{\cal B}_1, \oo\ell_{1};  \H\bigr),
\eaaa
and the space
$$
\V\defi \W\cap {\cal C},
$$
with the  norm $ \| u\|_{\V} \defi \| u\| _{{\W}} +\|
u\| _{{\cal C}}. $  Clearly, ${\cal C}^1\subset \V$.

\section{Problem setting and the main result}\label{SecM}

Let   $T>0$, $r\in \C$, and $\mu\in\HH$ be given. We consider the boundary value problem
 \baa
&&\frac{1}{i}\frac{d u}{d t}(t)=A u(t), \quad t\in (0,T),\label{eq}\\
&&\int_0^Te^{rt} u(t)dt=\mu.\label{ppp}\eaa
\par
We consider problem (\ref{eq})-(\ref{ppp}) assuming that  the initial value
$u(0)$  is unknown.

 \par
As usual, we accept that equation (\ref{eq}) is
satisfied for $u\in \V$ if,  for any $t\in[0,T]$,
\baa
\label{intur1} (u(t),\zeta)_H=(u(0),\zeta)_H+i\int_0^t (A u(s),\zeta)_Hds\quad
\forall \zeta\in\H.
\eaa
\begin{theorem}
\label{ThM} Assume that $\Re r\neq 0$. In this case, for any  $\mu\in \HH$, there exists a unique solution
$u\in\V^1$ of problem (\ref{eq})-(\ref{ppp}). This solution belongs to ${\cal C}^1$. Moreover, there exists $c>0$  such
that
\baa \|u\|_{{\cal C}^1}\le c\|\mu\|_{\HH}\label{estp} \eaa for all
$\mu\in \HH$. Here $c>0$ depends only on $\HBC,A,T$, and $r$. \end{theorem}
\par
By Theorem \ref{ThM}, problem
(\ref{eq})-(\ref{ppp}) is well-posed in the sense of Hadamard for $\mu\in\HH$.
\par
The proof of this theorem is given below; it is based on  construction of the solution $u$ for given $\mu$.

\section{Proofs}
\label{SecP}
Let us introduce  operators $\L :\H\to \V$  such that $\L\xi=u$,
where $u$ is the solution in $\V$ of  problem (\ref{eq}) with the Cauchy condition
\baa
u(0)=\xi\label{uxi}.
\eaa

Let a linear operator $M_0: H\to H $ be defined
such that \baaa
M_0 \xi=\int_0^T u(t)dt,\quad \hbox{where}\quad u=\L\xi.
\eaaa

In these notations, $\mu=M_0 u(0)$ for  a solution $u$ of problem  (\ref{eq}).

\begin{lemma}
\label{lemma1} Let us assume  that $\lambda_1\ge 1$ and that $c_A=0$.  Then the  linear operator $\L: \H\to {\cal C}^1$  is continuous, and  $\|\L\|\le 1$ for its norm.
\end{lemma}

It can be noted that, under the assumptions of Lemma  \ref{lemma1}, it follows from the definitions that
\baa
&&\|u\|_{\H}=-(u,Au)_H=\sum_{k=1}^{\infty}|\a_k|^2 \lambda_k,\quad u\in\H\nonumber\\
&&\|u\|_{\HH}=(Au,Au)_H=\sum_{k=1}^{\infty}|\a_k|^2 \lambda_k^2,\quad u\in\HH\label{HHH}.
\eaa

{\em Proof of Lemma \ref{lemma1}}.
 Let $\xi$  be expanded  as
\baa
\xi=\sum_{k=1}^\infty \a_k v_k.
\label{xi}\eaa
Here the coefficients $\a_k$ are such that $\sum_{k=1}^{+\infty}|\a_k|^2=\|\xi\|_H<+\infty$.
For $u=M_0\xi$, we have that \baa
u(t)=\sum_{k=1}^{\infty} \a_k e^{-i\lambda_k t}v_k.
\label{sol}\eaa
Clearly, we have that $\|u(t)\|_H\le \|\xi\|_H$ for all $t$.
In addition,
\baaa
-(u(t),Au(t))_H=\sum_{k=1}^{\infty} |\a_k|^2 \lambda_k e^{-i\lambda_k t}.
\eaaa
Hence, for all $t$,
\baaa
\|u(t)\|_\H=-(u(t),Au(t))_H\le \sum_{k=1}^{\infty} |\a_k|^2 \lambda_k=\|\xi\|_\H.
\eaaa
The continuity of $u(t)$ in $\H$ with respect to $t$ follows from
the Lebesgue's dominated converges theorem.   
This completes the proof of Lemma \ref{lemma1}. $\Box$

\begin{lemma}
\label{lemma2}  Under the assumption of Lemma \ref{lemma1}, the operators
 $M_0:\H\to \HH$ and 
 $M_0^{-1}:\HH\to \H$ are  continuous, and the statement of Theorem \ref{ThM} holds.
\end{lemma}

{\em Proof of Lemma \ref{lemma2}}.
  Let $\xi$  be expanded  as in (\ref{xi}), and let $\xi$  be expanded  as
\baa
\mu=\sum_{k=1}^\infty \g_k v_k.
\label{mu}\eaa

If $u\in\V $ is a solution of problem (\ref{eq})-(\ref{ppp}), then $u(0)\in H$
is uniquely defined; it follows from the definition of $\V $. Hence $\xi=u(0)\in H$
is uniquely defined. Let $\xi$ and $\mu$ be expanded  as
\baaa
\xi=\sum_{k=1}^\infty \a_k v_k,\quad \mu=\sum_{k=1}^\infty \g_k v_k,
\eaaa
 where
$\{\a_k\}_{k=1}^\infty$ and $\{\g_k\}_{k=1}^\infty$ and square-summable sequences in $\C$.
 By the choice of $\xi$, we have that  $u=\L \xi$. Applying the Fourier method, we obtain expansion (\ref{sol}).
\par
On the other hand,
\baaa
&&
\mu=\sum_{k=1}^{\infty} \g_k v_k(x)= \int_0^Te^{rt}u(t)dt=
\sum_{k=1}^{\infty} \int_0^Te^{rt}\a_k e^{-i\lambda_k t} v_k(x)dt =
\sum_{k=1}^{\infty} \zeta_k \a_k v_k(x),
\eaaa
where
\baa \zeta_k =\int_0^Te^{rt-i\lambda_k t}dt=\frac{1}{r-i\lambda_k}\left(e^{rT-i\lambda_k T}-1\right).
\label{zeta}\eaa
Hence \baaa
\|\mu\|_\HH^2 \le \sum_{k=1}^\infty \lambda_k^2\left|\frac{1}{r-i\lambda_k}\right|\left|e^{rT-i\lambda_k T}-1\right|^2|\a_k|^2 \le \w c\sum_{k=1}^\infty \lambda_k|\a_k|^2\le\w c \|\xi\|_\H^2,
\eaaa
where $\w c\defi \sup_k\left|e^{rT-i\lambda_k T}-1\right|^2$. This proves that the operator $M_0:\H\to\HH$ is continuous. 

Further,  the sequence $\{\a_k\}$ is uniquely defined as
\baa
\a_k =\g_k/\zeta_k,
\quad k=1,2,....
\label{ak}
\eaa
Since $\Re r\neq 0$, it follows that $\inf_k|\left(e^{rT-i\lambda_k T}-1\right)|\ge |e^{rT}-1|>0$.
Hence
\baaa
|\zeta_k^{-1}|\le \frac{\sqrt{|r|^2+\lambda_k^2}}{|e^{rT}-1|}.
\eaaa

It follows that
\baa
|\a_k|\le \frac{ \sqrt{|r|^2+\lambda_k^2}}{|e^{rT}-1|}|\g_k|
\eaa
and
\baaa
\sum_{k=1}^{\infty}\lambda_k |\a_k|^2 \le
\sum_{k=1}^{\infty}  \lambda_k |\g_k|^2\psi_k,
\eaaa
where
\baaa
\psi_k\defi \frac{1}{|e^{rT}-1|}\sqrt{|r|^2+\lambda_k^2}.
\eaaa
We have that
\baaa
|\psi_k|\le  \frac{\sqrt{r^2+\lambda_k^2}}{|e^{rT}-1|}\le \frac{ |r|+\lambda_k }{|e^{rT}-1|}.
\eaaa
By (\ref{HHH}), it follows that $\|u\|_{\H}\le \|u\|_{\HH}$ for $u\in\HH$. Hence
\baa
\|u(0)\|_\H=\|\xi\|_\H\le \frac{1}{|e^{rT}-1|}(\|\mu\|_\HH+|r| \|\mu\|_\H)\le  \frac{ 1+|r| }{|e^{rT}-1|} \|\mu\|_\HH.
\eaa
This proves that the operator $M_0^{-1}:\HH\to\H$ is continuous. 

The prior estimate (\ref{estp}) follows from Lemma \ref{lemma1}; the uniqueness
of solution follows from the linearity of the problem and from (\ref{estp}).
This complete  the proof of Lemma \ref{lemma2}. $\Box$

{\em Proof of Theorem \ref{ThM}}.
Under the assumption of Lemma \ref{lemma1},  the statement of  Theorem \ref{ThM} follows from Lemma \ref{lemma2}  and  Lemma \ref{lemma1}.
Let us consider the general case.

Let  $q>0$ be such that $\lambda_k+q\ge 1$ for all $k$.
Consider the eigenvalue  problem
\baaa
Au-qu=-\lambda u,\quad u\in \HBC.\eaaa
Clearly,  the eigenfunctions can
be selected the same as
for problem (\ref{EP}) with $q=0$, and the corresponding eigenvalues $\{\oo\lambda_k\}_{k=1}^\infty$ are
$\oo\lambda_k=\lambda_k+q\ge 1$.

Let $\oo r=r+iq$.  Clearly,  $\Re r\neq 0$ if and only if  $\Re\oo r\neq 0$.

Consider the boundary value problem
 \baa
&&\frac{1}{i}\frac{d\oo u}{d t}(t)=A\oo u(t)+q\oo u(t), \quad t\in (0,T),\label{eqq}\\
&&\int_0^Te^{\oo r t}\oo u(t)dt=\mu.\label{pppq}\eaa
By Lemma \ref{lemma2}, for any $\mu\in \HH$,
there exists a unique solution $\oo u(t)\in {\cal C}^1$ of this problem, and there exists a constant  $\oo c>0$ such
that
\baaa \|\oo u\|_{{\cal C}^1}\le \oo c\|\mu\|_{\HH}\label{estpq} \eaaa for all
$\mu\in \HH$. This $\oo c>0$ depends only on $\A,\H,T$, and $\oo r$.

Let \baa
u(t)\defi e^{i q t}\oo u(t).\label{uoou}
\eaa For  $t\in (0,T)$, we have that
\baaa
&&\frac{1}{i}\frac{d u}{d t}(t)=
\frac{1}{i}\frac{d\oo u}{d t}(t) e^{i q t}+ q \oo u (t)e^{i q t} = [A\oo u(t)-q\oo u(t)]e^{i q t}+q \oo u(t)e^{i q t}= Au(t) .\label{eqq2}\eaaa
In addition,
\baa
\int_0^Te^{r t} u(t)dt=\int_0^Te^{r t}\oo u(t)e^{i q t}dt
=\int_0^Te^{r+iqt }\oo u(t)dt=\int_0^Te^{\oo r t }\oo u(t)dt=\mu.\eaa
Hence 
$u\in{\cal C}^1$ is a  solution of problem (\ref{eq})-(\ref{ppp}). Estimate (\ref{estp}) follows from (\ref{estpq}) and from the choice of $u$.  To prove the uniqueness of the solution $u$, it sufficient to observe
that, for each $u\in\V^1$ satisfying (\ref{eq})-(\ref{ppp}), one can construct $\oo u\in \V^1$
 satisfying (\ref{eqq})-(\ref{pppq}) using  the steps described above in reverse.  This $\oo u$ is unique for a given $\mu$, and hence  $u$ is also unique.
 This completes the proof of Theorem \ref{ThM}. $\Box$
 
 \begin{remark}{\rm  Equations (\ref{sol}),(\ref{zeta}),(\ref{ak}), and (\ref{uoou}) imply an explicit eigenfunction expansion for the solution 
 \baa
u(t)=\sum_{k=1}^{\infty}  \frac{ r-i\lambda_k}{e^{rT-i\lambda_k T}-1} \g_k e^{-i\lambda_k t}v_k.
\label{u}
\eaa 
This can be used for numerical solutions.
  }
  \end{remark}
\begin{remark}{\rm It can be noted that Theorem 1 requires that $\mu$ belongs to $\HH$ rather than to $\H$, i.e.
it is more ”smooth” than the corresponding $u(0)$ that belongs to $\H$ only. In fact, this is quite
expectable: as it is seen from Lemma 2 and its proof, the integration over time is damping the
members of the eigenfunction expansion corresponding to larger $\lambda_k$. This explains why $\mu$ is more
”smooth” than $u(0)$.
}\end{remark}
\begin{remark}{\rm The regularity established in Theorem \ref{ThM} 
does not hold for an interesting case  where $r=0$. The choice of $r$ with $\Re r\neq 0$
was used to achieve regularisation. It can be noted that a similar regularity holds for   
a boundary value problem   
\baa
&&\frac{1}{i}\frac{d w}{d t}(t)=A w(t)-irw(t), \quad t\in (0,T),\label{eqw}\\
&&\int_0^Te^{rt} w(t)dt=\mu.\label{pppw}\eaa
for a Schr\"odinger equation with complex valued addition  $-r$ to the potential such that $\Re r\neq 0$.  In fact,  $w(t)=e^rtu(t)$.   }\end{remark}

\section{Conclusions}
The paper establishes solvability  of a boundary value problem 
for a   Schr\"odinger equations.  This problem
 can be reformulated as a new boundary value problem where a Cauchy condition is replaced by
 a prescribed time-average of the solution. It is shown that this new problem is well-posed in certain classes of solutions. This supplements existing results \cite{AS,BP,SM} for  
Schr\"odinger equation with non-local condition. These works  use the contraction mapping  theorem;  respectively, the non-local $Bu$ in time part of  the conditions such $u(0)-Bu=\mu$ 
has to be relatively small comparing with the term $u(0)$ representing the initial state of the solution in the condition.
Our approach is different and it does not require this restrictions since it  does not involve $u(0)$ explicitly. On the other hand, our approach is not applicable to nonlinear problems considered
in \cite{BP}.

So far, we have considered the case where the solutions can be expanded via basis from the eigenfunctions.  It would be interestingly to extend the result  on 
the more general case, as it was done in \cite{GG} for wave equations with periodic condition. 
We leave it for the future research.  

\end{document}